\documentclass[11pt]{article}
\newcommand{\ieeeonly}[1]{}
\newcommand{\lncsonly}[1]{}
\newcommand{\articleonly}[1]{#1}
\newcommand{\acmonly}[1]{}

\ieeeonly{\IEEEoverridecommandlockouts} %

\newcommand{\myparagraph}[1]{\paragraph{#1.}}

\newcommand{\mysection}[1]{\section{#1}} %

\newcommand{\short}[1]{} %
\newcommand{\shortmed}[1]{#1} %
\newcommand{\medium}[1]{#1} %
\newcommand{\medfull}[1]{#1} %
\newcommand{\full}[1]{} %
\newcommand{\unused}[1]{} %

\lncsonly{\pagestyle{headings}}
\acmonly{\settopmatter{printfolios=true}}

\acmonly{
\settopmatter{printacmref=true}
\fancyhead{}

\def\BibTeX{{\rm B\kern-.05em{\sc i\kern-.025em b}\kern-.08emT\kern-.1667em\lower.7ex\hbox{E}\kern-.125emX}}
}

\ieeeonly{\usepackage[subtle,mathspacing=normal]{savetrees}} %

\usepackage{hyperref}
\usepackage{enumitem}  %
\usepackage{outlines}
\setlist{itemsep=0em}
\usepackage{graphicx}
\full{\usepackage{multirow}}
\usepackage{xspace}
\articleonly{\usepackage{amssymb}}\ieeeonly{\usepackage{amssymb}}
\usepackage{floatflt}

\articleonly{
\usepackage{authblk}
\usepackage{setspace}
\setstretch{1.07}
\usepackage{fullpage} %
}

\ieeeonly{\newtheorem{lemma}{Lemma}
\newtheorem{theorem}{Theorem}}
\articleonly{\newtheorem{lemma}{Lemma}
\newtheorem{theorem}{Theorem}}

\newcommand{\tuple}[1]{\langle #1\rangle}

\newcommand{\setc}[2]{\{#1 \;|\; #2\}}
\newcommand{\intersect}{\cap}
\newcommand{\union}{\cup}
\articleonly{\newcommand{\implies}{\Rightarrow}}\ieeeonly{\newcommand{\implies}{\Rightarrow}}\lncsonly{\newcommand{\implies}{\Rightarrow}}
\def\mathify#1{\ifmmode{#1}\else\mbox{$#1$}\fi}

\newcommand{\var}[1]{\mathify{\mathit{#1}}}
\newcommand{\func}[1]{\mathify{\mathit{#1}}}
\newcommand{\class}[1]{\mathify{\mathit{#1}}}
\newcommand{\const}[1]{\mathify{\mathsf{#1}}}  %

\newcommand{\RepProt}{{RepProt}}
\newcommand{\repProt}{{repProt}}
\newcommand{\SI}{SI}
\newcommand{\RSI}{RSI}
\newcommand{\rsence}{RS-equivalence\xspace}
\newcommand{\rsent}{RS-equivalent\xspace}
\newcommand{\reloc}{\func{reloc}}
\newcommand{\fmlanote}[1]{(#1)}
\ieeeonly{\newcommand{\qed}{$\Box$}}
\articleonly{\newcommand{\qed}{$\Box$}}

\begin{document}

\acmonly{
\setcopyright{none}
\fancyhead{}
}

\newcommand{\thanksText}{This material is based on work supported in part by 
    ONR grants N000142112719 %
    and N000142012751 %
    and NSF grant CCF-1954837. %
     }

\title{Resilience through Automated Adaptive Configuration for Distribution and Replication%
\ieeeonly{\thanks{\thanksText}}\lncsonly{\thanks{\thanksText}}
}

\ieeeonly{\author{
\IEEEauthorblockN{Anonymous}
}}

\articleonly{\author{Scott D.~Stoller ~~~~~~~ Balaji Jayasankar ~~~~~~~ Yanhong A.~Liu}
  \affil{Department of Computer Science, Stony Brook University, Stony Brook, NY, USA}}

\lncsonly{\author{Scott D.\ Stoller and Balaji Jayasankar and Yanhong A.\ Liu}
  \institute{Department of Computer Science, Stony Brook University, Stony Brook, NY, USA}}

\acmonly{
\author{Scott~D.~Stoller}
\affiliation{\institution{Stony Brook University}\country{U.S.A.}}
\email{stoller@cs.stonybrook.edu}
\author{Balaji Jayasankar}
\affiliation{\institution{Stony Brook University}\country{U.S.A.}}
\email{bjayasankar@cs.stonybrook.edu}
\author{Yanhong A.~Liu}
\affiliation{\institution{Stony Brook University}\country{U.S.A.}}
\email{liu@cs.stonybrook.edu}
}

\newcommand{\abstracttext}{ %
This paper presents a powerful automated framework for making complex systems resilient under failures, by optimized adaptive distribution and replication of interdependent software components across heterogeneous hardware components with widely varying capabilities.  A {\em configuration} specifies how software is distributed and replicated: which software components to run on each computer, which software components to replicate, which replication protocols to use, etc.  We present an algorithm that, given a system model and resilience requirements, (1) determines initial configurations of the system that are resilient, and (2) generates a reconfiguration policy that determines reconfiguration actions to execute in response to failures and recoveries.  This model-finding algorithm is based on state-space exploration and incorporates powerful optimizations, including a quotient reduction based on a novel equivalence relation between states.  We present experimental results from successfully applying a prototype implementation of our framework to a model of an autonomous driving system.
}

\acmonly{
\begin{abstract}
\abstracttext
\end{abstract}}

\acmonly{}

\maketitle
\ieeeonly{
\thispagestyle{plain}\pagestyle{plain}
\begin{abstract}
\abstracttext
\end{abstract}}
\lncsonly{
\begin{abstract}
\abstracttext
\end{abstract}}
\articleonly{
\begin{abstract}
\abstracttext
\end{abstract}}

\mysection{Introduction}
\label{sec:intro}

Increasingly sophisticated critical software systems are built from a myriad of interconnected components providing a multitude of functionalities and running on diverse hardware components.
It is well-known that distributing and replicating functionalities can enhance reliability.
Reliability can be further enhanced by making a system \emph{resilient}, i.e., able to quickly adapt to failures in order to continue providing service.
To achieve resilience with minimum cost, systems must be able to adapt, to optimally utilize the available hardware resources after any sequence of hardware failures and recoveries.

This paper presents a powerful automated framework for making complex systems resilient under failures, by optimized adaptive distribution and replication of software components across heterogeneous hardware components with widely varying capabilities, ranging from compute servers to mobile devices.  Key features of the framework, and the main contributions of this paper, are:

\begin{itemize}
\item \emph{Precise system model}, based on key characteristics needed for automated optimized adaptive distribution and replication, covering hardware, software, replication protocols, and configuration, and including dependence, compatibility, and capacity constraints.  A \emph{configuration} specifies how software is distributed and replicated: which software components run on each computer, which software components are replicated, the number of replicas and replication protocol used for each software, etc.

Identifying these characteristics, thereby enabling automation of these decisions for a large class of systems, is a key contribution.
The model supports a variety of hardware device types, because the challenge we address is especially important in cyber-physical systems. 

\item \emph{Resilience requirements}, specifying the \emph{failure model}---the types of failures, as well as their numbers and rates, that must be tolerated---and the \emph{critical functionalities}---functionalities that must be continuously available.  The limit on failure rate helps ensure that the system has the opportunity to reconfigure in response to some failures, before additional failures occur.\short{  A few illustrative scenarios appear in Section \ref{sec-reconfig}.}

We focus on hardware failures, because
software failures are addressed using, e.g., retry blocks and N-version programming, not distribution and replication.

\item \emph{Adaptive configuration policy generator}, in the form of a model-finding algorithm that, given a system model and resilience requirements, solves two core problems: (1) determine \emph{initial configurations}---configurations before any failures occur---that are \emph{resilient}, i.e., guarantee continuous availability of all critical functionalities if the system starts in that state and appropriate reconfiguration actions are taken after each failure and recovery; and (2) generate a \emph{reconfiguration policy} that determines reconfiguration actions to execute in response to each failure and recovery.

Reconfiguration actions include adding and removing software replicas; starting, stopping, and moving running software; and replacing running software with alternative software with the necessary functionalities but different hardware requirements, software dependencies, or QoS.  Precise modeling of replication protocols ensures that the best one is used in each context. 

\item \emph{Quotient reduction based on a novel equivalence relation between states}, and other optimizations, to increase scalability of state-space exploration performed by the configuration policy generator.

\item \emph{Prototype implementation and experimental results} from successfully applying it to a model of an autonomous driving system. 
\end{itemize}

\medfull{To illustrate the benefit of reconfiguration, consider a service replicated using primary-backup replication in a system with 3 computers and whose failure model specifies that there are at most 2 concurrently failed computers at any time, and that the time between consecutive failures is larger than the system reconfiguration time.  To ensure continuous availability of the service without reconfiguration, all three computers must be replicas.  With reconfiguration, it is sufficient to have two live replicas when zero or one computer is dead, and to have one live replica when two computers are dead, provided that after each failure of a replica, the system reconfigures to remove the dead replica and, if there is a live computer that isn't a replica, add it as a new replica.  Section \ref{sec-reconfig} presents additional examples of situations where reconfiguration is beneficial.}

We implemented the entire framework, with explicit-state state-space exploration, in DistAlgo~\cite{Liu+17DistPL-TOPLAS,distalgo22git}, an extension of Python with support for logic quantifications with patterns as queries (and for high-level distributed programming, which would be useful later to implement a runtime system for reconfiguration).  This allows the mathematical definitions of reconfiguration, resilience, etc. to be translated into code directly and easily.

We also implemented most of the framework using the SMT solver Z3~\cite{z3smt2008}, with a configuration policy generator based on symbolic state-space exploration.  We encountered multiple difficulties, including the need to use\medfull{ low-level} Boolean encodings of our models and constraints due to lack of direct support for quantification over sets and functions, lack of direct support for recursion, and limited support for optimization problems\medfull{, as discussed in Section \ref{sec-smt}}.

\short{Many details, including further discussion of our Z3 implementation, are omitted due to space limitations and are available in~\cite{adaptiveDist-arxiv}.}  %

\mysection{Related work}
\label{sec:related}

Many works on replication protocols refer to support for reconfiguration.  This means that the protocol supports dynamically adding and removing replicas, not that it addresses the system configuration problems we tackle.

There is a sizable literature on self-adaptive systems, surveyed in~\cite{schneider2015,weyns2020,wong2022}.  Weyns~\cite{weyns2020} identifies four essential tasks for self-adaptation.  Our work is primarily self-healing and self-configuration---plus some overlap with self-protection and self-optimization---in the context of distribution (placement) and replication of computations.  Work on self-adaptive systems addresses these tasks, especially self-healing, mainly in the context of other aspects of system configuration, such as network routing.  Self-adaptation of these other aspects of configuration is complementary to our work.

Several works address configuration of distribution and replication of computations in specific settings. 
For example, 
Daidone et al.~\cite{daidone2013} use stochastic simulations to evaluate how configuration parameters, e.g., heartbeat rate for failure detection, of a single service replicated on a homogeneous set of servers affect the system's performance.
Machida et al.~\cite{machida2010} optimize static placement of VMs providing independent services on servers with different capacities, to reduce the number of servers needed to tolerate a given number of crashes.  
Stoiescu et al.~\cite{stoicescu2017} design a software architecture that supports adding and removing software components that implement fault-tolerance mechanisms, such as replication and retry blocks, at runtime.  
{\sc Hamraz}~\cite{li2022} is a programming language\medfull{ with a sophisticated type system,} whose compiler decides where to replicate objects and method invocations to ensure specified availability (and other) properties in distributed systems with multiple administrative domains and subject to Byzantine attacks.
Our work has two novel features compared to existing work.  First, our definition and analysis of resilience are recursive.  This enables more precise analysis of resilience, compared with traditional non-recursive definitions, for systems that reconfigure in response to failures, because it checks, for each sequence of failures, whether the system configuration created in response to those specific failures can tolerate exactly the failures that can occur in that failure state, based on the assumptions in the failure model.
Furthermore, our analysis synthesizes configurations and reconfiguration policies and provides precise resilience guarantees for them in advance, rather than simply making a ``best effort'' in each state without advance guarantees. \medfull{

}Second, we consider automatic selection of replication protocols.  Jointly optimizing the choice of replication protocol and the number and placement of replicas for each software component, and making them adaptive (varying over time), enables our method to achieve higher resilience with given resources than prior work.   Also, our work uses a more comprehensive system model than much prior work, taking software dependencies, hardware compatibility, resource constraints, etc., into account. \medfull{

}
Overall, our work is novel in providing these features and generating adaptive configuration policies that are guaranteed to satisfy resilience requirements.

\medfull{
There is work on software (re)configuration for resilience not in the context of distributed systems.
For example, Strunk and Knight~\cite{strunk2006} present a high-level system architecture that supports software reconfiguration and an approach to formally verifying that parameters of software modules are reconfigured correctly.
Wu et al.~\cite{wu2012} present a software-architecture approach to statically extend a system with software components implementing fault-tolerance mechanisms, such as retry blocks and N-version programming.
These techniques are complementary to our work.
}

\medfull{Bloem et al.~\cite{bloem2024} model security-related aspects of systems and use MaxSMT to find minimum-cost changes that protect from security threats.  The approach is similar to our use of Z3\short{ (see Section \ref{sec:intro})}, but their problem is simpler: it does not consider adaptation, does not involve recursion, etc.}

\mysection{System model}
\label{sec-system}

Our object-oriented system model captures characteristics of hardware components, software components, and replication protocols that affect adaptive configuration for resilience, including dependence, compatibility, capacity, and distribution constraints.
\medfull{ 
 Every object has an attribute \var{class} (the object's type).  We write long conjunctions and disjunctions as itemized lists bulleted with $\land$ and $\lor$, respectively~\cite{lamport94how}.}

\subsection{Hardware}

Class \class{Computer} is used to model devices that can run application software: embedded computers, laptops, cell phones, etc. 
\short{It has attributes specifying the: OS and CPU architecture; computational resources, namely, cores and RAM; networking capabilities (wired, wifi, and/or cellular); power sources, which may be internal (e.g., battery) or external; integrated devices (i.e., devices built into the computer), represented as a set \var{devices} of types of integrated devices, e.g., \const{GPS} or \const{camera}.}%
\medfull{It has attributes specifying the: operating system \var{OS}; CPU architecture \var{CPUarch}; 
amount of RAM \var{RAM} and number of cores \var{cores}; set of types of integrated devices \var{devices}, e.g., \const{GPS} or \const{camera}; networking capabilities, specified by Booleans \var{wiredNIC}, \var{wifiNIC}, and \var{cellular}; and set of power sources \var{power}, which may be internal, e.g., a battery, or external, e.g., a UPS.  We distinguish wired and wifi NICs because some applications may require the former's higher performance.}
We model power sources, because a power source can be a single point of failure for multiple devices.
 
 \short{Class \class{Device} is used to model stand-alone devices, i.e., devices not integrated into computers.  It has attributes specifying the: device type \var{type}, e.g., \const{camera}; and power sources (none for devices  (e.g., power sources) that do not require a power source).}%
 \medfull{Class \class{Device} is used to model stand-alone devices, i.e., devices not integrated into computers.  It has attributes: \var{type}, device type, e.g., \const{camera}, and \var{power} (same meaning as for \func{Computer}; $\emptyset$ if the device does not require a power source, e.g., if the device is itself a power source).}
Stand-alone devices are assumed to be network-connected and accessible by all computers.

We refer to  objects of type \class{Computer} or \class{Device} as ``hardware components''.
Network hardware, such as routers and wireless access points, is not currently included in our model.
\full{We assume wifi and wireless networks are used for communication between components of the system, and cellular connections are used to communicate with external resources.  Furthermore, we assume there is reliable communication between all computers and devices with wired or wifi NICs.} 

\subsection{Software}

Our model considers dependencies between software components as well as dependencies on hardware.  This is especially useful for systems in which larger software components have been decomposed into smaller ones, which can be distributed and replicated differently to make the best use of limited resources.  The model captures characteristics that determine whether a software component must be replicated to achieve resilience or whether starting a new instance of the software after a failure is sufficient.

\short{The \class{Software} class has attributes specifying the: functionality \var{fn}, e.g., \const{map} \const{database} or \const{planning}; set of required functionalities \var{fnReq}, i.e., functionalities provided by other software and used by this software; set \var{devices} of types of required devices, e.g., \const{camera} for vision software; required OS and CPU architecture, if any; resource requirements, namely, cores and RAM; and required networking capabilities (wired, wifi, and/or cellular).}%
\medfull{The \class{Software} class has attributes specifying the: functionality \var{fn}, e.g., \const{map} \const{database} or \const{planning}; set of required functionalities \var{fnReq}, i.e., functionalities provided by other software and used by this software; set of types of required devices \var{devices}, e.g., \const{camera} for vision software; required CPU architecture \var{CPUarch} if any (\const{None} otherwise); required operating system \var{OS} if any (\const{None} otherwise); required amount of RAM \var{RAM} and number of cores \var{cores}; and required networking capabilities, specified by Booleans \var{cellular} and \var{wired} (wireless is assumed sufficient if both are false).}
It also has the following Boolean attributes specialized to resilience analysis:
\begin{itemize}
\item \var{deterministic}: whether the software is deterministic, i.e., different processes running it and that receive the same sequence of inputs produce the same outputs.  This affects which replication protocols can be used.
\item \var{fastStarting}: whether the software's startup time is less than or equal to the acceptable downtime of its functionality.  If not, replication, rather than starting new instances, must be used to achieve continuous availability of this software.
\item \var{migratable}: whether a running instance of the software can be moved to another computer.  This depends on capabilities of the OS as well as characteristics of the software: support for migration of essential volatile state (if any), redirection of network connections (if any), etc.
\item \var{persisState}: whether the software maintains essential mutable persistent state.\medfull{  Note that this is independent of whether the software asks other software to store persistent state on its behalf.}
\item \var{preferred}: whether this software is preferred, e.g., provides higher QoS than non-preferred software with the same functionality.  This supports generating reconfiguration policies that achieve graceful degradation of QoS due to failures.
\full{This Boolean could easily be replaced with a quantitative preference specification.}
\item \var{remoteUse}: whether the software can be used remotely by software components running on other computers.\medfull{  This affects which software components need to be co-located.}
\item \var{resumable}: whether the software functions correctly after the computer it is running on fails and then recovers.  This holds if the software does not have essential volatile state\unused{ (i.e. volatile state that, if lost, causes unacceptable disruption to the functionality)} and can resume functioning correctly if it misses events while it is down.\full{  For example, scene understanding components and file servers may be resumable.}
\item \var{singleInstance}: whether there should be at most one (possibly replicated) instance of this software in a system.  This generally holds if it maintains data that should be unique system-wide.  %
For example, a vehicle's computer vision system might run multiple independent object recognition components (one per camera) and feed the results into a single-instance scene understanding component that creates an integrated model of the vehicle's surroundings.\medfull{  For simplicity, we require that software with persistent state be single-instance.  This ensures that the persistent state is uniquely identified by the software component that created it.}\full{  Without this requirement, we would need to extend the definition of configurations to keep track of which instance of a software component created some persistent state (e.g., ``the instance that was initially running on computer c1 and later moved to c2'').\unused{  This requirement is consistent with our assumption that a software component's identity expresses aspects of its configuration, since the configuration often identifies a distinctive (unique) purpose for the software component, e.g., ``analyze data from camera X and periodically save selected results''.}  Note that this requirement constrains how systems are modeled, but does not limit their designs.  It can always be satisfied by including more information in a software component's identity or functionality, to distinguish different instances.}
\item \var{smallPersisState}: whether the persistent state (if any) is small, in the sense that the time needed to copy it over the network during reconfiguration is acceptable.\full{  This attribute is meaningful only if persisState=true.} 
\end{itemize}

\subsection{Knowledge base for replication protocols}

Our model includes a knowledge base for replication protocols, capturing key characteristics needed to support automated selection of the best applicable protocol in each context.  The \var{\RepProt} class has attributes: \var{sync}, a Boolean indicating whether the protocol is only for synchronous systems; \var{failTypes}, the types of handled failures; $\var{active}$, a Boolean indicating whether active or passive replication is used; \var{progressQ}, an enum with allowed values \const{majority} and \const{all}, indicating the size of a quorum needed for successful execution of a client request; and \var{reconfigQ}, an enum with allowed values \const{majority} and \const{one}, indicating the size of a quorum needed for successful reconfiguration, a.k.a.\ membership change.  For brevity, we omit generic attributes of software components, such as required OS; they can trivially be added.\unused{  We plan to extend this initial knowledge base with additional attributes: persistent storage requirements, failover time (for protocols with a leader or primary), etc.}

In active replication, all replicas execute requests from clients; this is suitable only for deterministic applications.  In passive replication, one replica, called the \emph{active replica} or \emph{primary}, executes requests from clients and disseminates the resulting state changes and outputs to the other replicas.\full{  This distinction affects the definition of available functionality (below): the functionality of an actively-replicated software component \var{sw} is available only if the devices and functionality it requires are available on a quorum of replicas; for a passively-replicated software component, they need to be available only on the primary.  The definition of valid configuration requires that all replicas have sufficient RAM and CPU to run the application, regardless of whether the replication is active or passive, since non-primaries typically run the application in a shadow mode, which continuously applies state updates received from the primary, so each non-primary can quickly take over as primary if needed.}

For traditional primary-backup protocols and chain replication protocols, $\var{progressQ} = \const{all}$ and $\var{reconfigQ} = \const{one}$.  For asynchronous state-machine replication protocols, such as Paxos~\cite{Lam98paxos,lamport2010reconfiguring}\short{ and}\medfull{,} Raft \cite{ongaro14raft}\medfull{ and Viewstamped Replication \cite{oki88vsr}}, typically $\var{progressQ} = \const{majority}$ and $\var{reconfigQ} = \const{majority}$.

A recovered replica can resume participating in a replication protocol without reconfiguration, provided it was not removed from the membership while it was down.

\subsection{Systems and configurations}
\label{sec:config}

\newcommand{\classdiagram}{
\begin{floatingfigure}[r]{.55\textwidth}
\includegraphics[width=.5\textwidth]{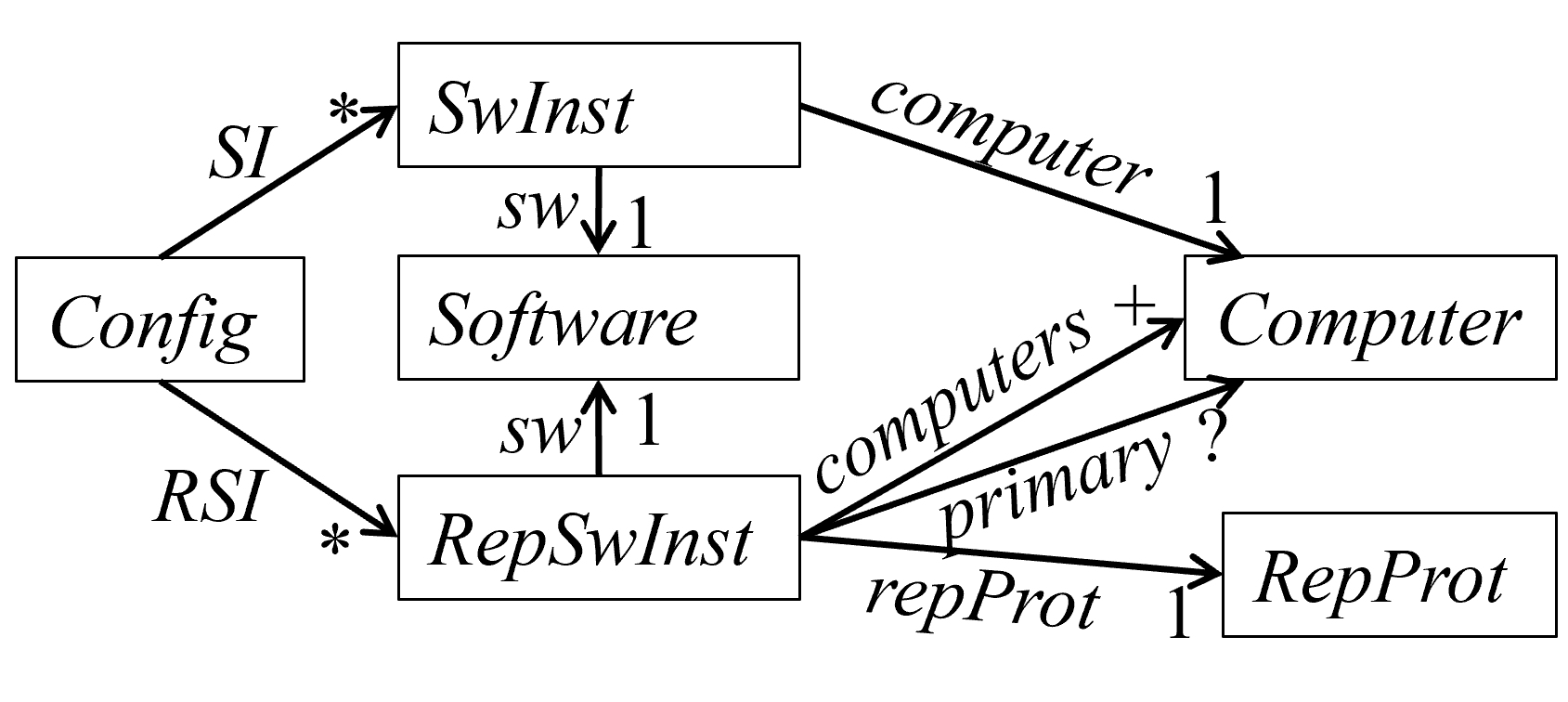}
\end{floatingfigure}}

\short{\classdiagram}

The \class{System} class has attributes specifying the sets \var{hw} and \var{sw} of hardware and software components, respectively, and a Boolean \var{sync} indicating whether the system is synchronous.\full{  Synchrony affects which replication protocols can be used.  It is straightforward to generalize this definition to allow different synchrony assumptions for different parts of a system.}\medfull{  Some definitions below are implicitly parameterized by the system $\var{sys}$; for example, ``there exists a computer $c$'' means ``there exists $c$ in $\var{sys.hw}$ with type $\var{Computer}$''.} \medfull{ \classdiagram

}A \emph{configuration} of a system specifies which software components run on each computing device, which software components to replicate, and, for each replicated software component, which replication protocol to use to coordinate the replicas.\medfull{  A {\em software instance} is a copy of some software running on either one computer, for an unreplicated software instance, or on multiple computers, for a replicated software instance.  Concretely, a software instance represents one or more processes running that software.  Although ``software instance'' could be interpreted to mean ``an object of type \class{Software}'', we never use this term in that sense; we refer to such an object as ``a software component''.}  The inlined UML class diagram shows classes \class{Config}, \class{SwInst} (for unreplicated software instances), \class{RepSwInst} (for replicated software instances), etc.  Multiplicity symbols 1, ?, *, and + mean ``one'', ``zero or one'' (a.k.a. ``optional''), ``zero or more'', and ``one or more'', respectively.  For example, the diagram shows that\medfull{ the attributes of \class{Config} includes sets of unreplicated and replicated software instances \var{SI} and \var{RSI}, respectively, and} the attributes of \class{RepSwInst} include a set of computers \var{computers} where replicas run, and an optional primary replica \var{primary}.
For an instance \var{rsi} of \class{RepSwInst}, we sometimes treat \var{rsi.\repProt.processQ} and \var{rsi.\repProt.reconfigQ} as numbers; for example, if $\var{rsi.\repProt.processQ} = \const{majority}$, then \var{rsi.\repProt.processQ} is interpreted as $\lceil (|\var{rsi.computers}|+1)/2 \rceil$.

\medfull{We allow replication only for \emph{stateful} software, i.e., software components \var{sw} for which \var{sw.persisState} is true or \var{sw.resumable} is false (indicating the presence of essential persistent or volatile state).\footnote{For economy of modeling, we use \var{resumable} to check for volatile state; a dedicated attribute for that could easily be introduced, if desired.}  This is reasonable because replication is used to replicate an application's state to avoid losing it\full{ (or losing access to it)} due to failures.}\full{  This requirement allows some simplification of the definition of reconfiguration, but could easily be eliminated if desired.}\medfull{  For simplicity, we allow at most one replicated software instance\medfull{ (replicated on any number of computers)} for each software component; this restriction can easily be eliminated by adding an id attribute to the \class{RepSwInst} class.}\unused{  This restriction is reasonable, because typically multiple instances of a software component are needed only if they are local to a particular computer.  Relaxing this restriction requires adding a unique identifier for each replicated software instance; for unreplicated software instances, the computer is sufficient to identify the instance, but for $\var{\RSI}$, the set of computers is not suitable for this, because it can change during reconfiguration.}

Our model captures several types of constraints on configurations.  A configuration \var{cfg} is {\em valid} if it satisfies these constraints.
{\em Dependence constraints} include dependence of software on software---for example, a software component that requires a functionality  provided by another---and of software on hardware---for example, a software component that requires data from a GPS.  
{\em Compatibility constraints} include (1) hardware/software compatibility, e.g., between binary executables and CPU architectures such as ARM and x86, and between a system's synchrony guarantee and a replication protocol's synchrony requirements, and (2) software/software compatibility, e.g., between application software and OS.
{\em Capacity constraints} ensure software resource requirements do not exceed hardware resource capacities: total number of CPU cores and amount of memory needed by the software components running on each computing device; capacity constraints for other resources, e.g., network throughput, can easily be added. %
{\em Distribution constraints} require co-location of software components---specifically, if a software component depends on functionality of another software component accessible only through a local API---and ensure there is at most one instance of each \var{singleInstance} software component.

\medfull{\bigskip}
\mysection{Failure model}
\label{sec-failure}

A {\em failure model} specifies the types of hardware failures considered, which types of hardware are subject to each type of failure, and the maximum number and rate of failures.  We consider crash failures and recovery from crashes in this paper.  
We assume detection of crash failures is complete, i.e., every failed component is eventually suspected of failure.  In synchronous systems, we further assume that it is accurate, i.e., correct components are not suspected of failure.
\medfull{Failure detection can be implemented using common techniques, such as heartbeat messages or ping-pong messages~\cite[section 8.2.8]{vansteen2023}\full{\cite[section 13.5]{birman2005reliable}}, or more sophisticated techniques based on probabilistic analysis, e.g., \cite{chen2002b}.}\full{  Failure detectors are implicit in our framework.  We do not model them as explicit software components, because they have modest hardware resource requirements and do not have specific hardware or software dependencies.}

A {\em failure} is modeled with class \class{Failure}.  It has attributes specifying the hardware component \var{hw} that failed and the failure type \var{type}, currently only \const{crash}.  A {\em recovery} from a failure is modeled in the same way as a failure; it represents the end of that failure. 

A {\em failed set} is the set of failures representing the currently failed hardware components with their failure types, i.e., the failures that have occurred minus the recoveries.
A \emph{state} of a system is a pair $\tuple{\var{cfg}, \var{fs}}$, where \var{cfg} is a configuration, and \var{fs} is a failed set.

\shortmed{
A hardware component \var{h} is \emph{live} with failed set $\var{fs}$, denoted $\func{live}(\var{h},\var{fs})$, if $h$ is not in $\var{fs}$, and, if $h$ requires an external power supply, at least one of its power supplies is live.
Functionality \var{fn} is \emph{available} on computer \var{c} in configuration \var{cfg} with failed set \var{fs}, denoted \func{availOn}(\var{fn},\var{c},\var{cfg},\var{fs}), if: either 
(1) \var{cfg} contains a software instance \var{si} running on a live computer that 
(1a) \var{si.sw} has functionality \var{fn}, 
(1b) if $\neg\var{si.sw.remoteUse}$, then $\var{c}=\var{si.computer}$, 
(1c) for each functionality \var{fn'} required by \var{si.sw}, \func{availOn}(\var{fn'},\var{c},\var{cfg},\var{fs}), and
(1d) each device type \var{d} required by \var{si.sw} is available on \var{c}, i.e., \var{c} has an integrated device of type \var{d} or there is a live network-accessible device of type \var{d}; or
(2) \var{cfg} contains a replicated software instance \var{rsi}, and there exists a subset \var{Q} of \var{rsi.computers}, such that 
(2a) \var{rsi.sw} has functionality \var{fn}, 
(2b) $|\var{Q}| \geq \var{rsi}.\repProt.progressQ$ and $\forall \var{c} \in \var{Q}: \func{live}(c,\var{fs})$,
(2c) if $\neg\var{rsi.sw.remoteUse}$ then $\var{c} \in \var{Q}$,
(2d) if $\var{rsi.\repProt.active}$, then 
$\forall \var{c'} \in \var{Q}: 
\func{availOn}(\var{si.sw.fnReq},\linebreak[0]\var{c'},\var{cfg},\var{fs})$ and 
each device type \var{d} required by \var{si.sw} is available on \var{c},
and (2e) if $\neg\var{rsi.\repProt.active}$, then $\func{availOn}(\var{si.sw.fnReq},\var{rsi.primary},\var{cfg},\var{fs})$ and 
each device type \var{d} required by \var{si.sw} is available on \var{rsi.primary}.
Functionality \var{fn} is \emph{available} in state $\tuple{\var{cfg}, \var{fs}}$,\medium{ denoted $\func{avail}(\var{fn},\tuple{\var{cfg},\var{fs}})$,} 
if it is available on some computer in that state.
}

A \emph{failure model} characterizes the sequences of failures  a system is expected to tolerate.  A failure model specifies (1) the maximum numbers of failed components, for various failure types and hardware types, and (2) the rate of failures, abstracted as the maximum number of ``simultaneous'' failures, i.e., the maximum number of failures that can occur in a time interval shorter than the system's reconfiguration time.

The \class{FailBound} class is used to specify limits on failures of a given type.  It has attributes specifying the hardware type  \var{hwType}, failure type \var{fType}, maximum number \var{n} of hardware components of \var{hwType} that can have a failure of type \var{fType} in any state, and maximum number \var{maxSimult} of simultaneous failures of type \var{fType} for hardware components of type \var{hwType} (\const{None} if there is no limit other than \var{n}).

The \class{FailureModel} class has attributes \var{fBounds} (a set of instances of \class{FailBound}) and \var{maxSimult} (the maximum number of simultaneous failures across all failure types, or \const{None} if there is no limit beyond the one implied by \var{fBounds}).

These bounds could be based on a probabilistic failure model and probabilistic reliability requirement, e.g., the expected time until a combination of failures that exceeds these bounds is at least $k$ years.  The bound on simultaneous failures should also take common-cause failures into account.

\full{A failure model \var{fm} is \emph{unlimited-rate} if all permitted failures can occur simultaneously, i.e., if $\var{fm.maxSimult} = \const{None}$ and $\var{fm.fBounds.maxSimult} = \{\const{None}\}$.}

A failed set \var{fs} is \emph{consistent} with failure model \var{fm}, denoted \func{consistent}(\var{fs}, \var{fm}), if\short{ it satisfies the bounds on the total number of failures and the numbers of failures of each type.}\medfull{
\begin{outline}
    \1[$\land$] $\forall \var{f} \in \var{fs}: \exists \var{fb} \in \var{fm.fBounds}: \var{f.hw.type} = \var{fb.hwType} \land \var{f.type} = \var{fb.fType}$
    \1[$\land$] $\forall \var{fb} \in \var{fm.fBounds}:
   |\{ \var{f} \in \var{fs} \;:\; \var{f.hw.type} = \var{fb.hwType} \land \var{f.type}=\var{fb.fType}\}| \leq \var{fb.n}$
\end{outline}}

Given a failed set \var{fs} and a failure model \var{fm}, let \func{nextFS}(\var{fm},\var{fs}) be the set of failed sets that are consistent with \var{fm}, are strict supersets of \var{fs}, and are reachable from \var{fs} by one set of simultaneous failures\medfull{, i.e., a set of failures that satisfies all \var{maxSimult} constraints}.\full{  This definition requires that the resulting failed sets are strict supersets of \var{fs} because it considers only failures, not recoveries.}
Let \func{nextFSworst}(\var{fm},\var{fs}) contain the subset-maximal members of \func{nextFS}(\var{fm},\var{fs}).\full{  This is equivalent to taking all of the largest members of \func{nextFS}(\var{fm},\var{fs}) (i.e., all \var{fs'} in \func{nextFS}(\var{fm},\var{fs}) with $|\var{fs'}| = \func{max}(\{|\var{fs''}| \;:\; \var{fs''} \in \func{nextFS}(\var{fm},\var{fs})\})$.}  When a maximal set of simultaneous failures occurs in a state with failed set  \var{fs}, one of these ``worst-case'' failed sets is reached.

\mysection{Reconfiguration model}
\label{sec-reconfig}

After a failure or recovery, the system may change its configuration using the following types of \emph{reconfiguration actions}:
\begin{itemize}
    \item \func{changeReps}(\var{rsi}, \var{C}, \var{c}): for replicated software instance \var{rsi}, change \var{rsi.computers} to set \var{C} of computers and, if \var{rsi} uses a passive replication protocol, change \var{rsi.primary} to computer \var{c}.
    \item \func{stop}(\var{si}): stop running software instance \var{si}
    \item \func{stopRep}(\var{rsi}): stop running replicated software instance \var{rsi}
    \item \func{start}(\var{si}): create and start running software instance \var{si}
    \item \func{move}(\var{si}, \var{c}): move a migratable software instance \var{si} to computer \var{c}, by migrating its volatile and persistent state, redirecting its network connections, etc.
\end{itemize}

\full{There is no action to create and start running a new replicated instance, for reasons discussed below.}

\medfull{We define a transition relation that expresses the preconditions and effects of each reconfiguration action.  It has the form $\tuple{\var{cfg},\var{fs}} \stackrel{\var{a}}\rightarrow \tuple{\var{cfg'},\var{fs}}$ where \var{a} is a reconfiguration action.  We first define an auxiliary predicate \func{canRun}(\var{c},\var{sw},\var{cfg},\var{fs}) that holds if computer \var{c} satisfies the requirements to run software \var{sw} in configuration \var{cfg} with failed set \var{fs}: $c$ is compatible with the software, and all of the necessary functionality, devices, and computational resources are available at \var{c}.\full{  Formally, 
\func{canRun}(\var{c},\var{sw},\var{cfg},\var{fs}) holds if:
\begin{outline}
    \1[$\land$] \func{compatible}(\var{sw}, \var{c})
    \1[$\land$] $\forall \var{fn} \in \var{sw.fnReq}: \func{availOn}(\var{fn},\var{c},\var{cfg},\var{fs})$,
    \1[$\land$] $\forall \var{dr} \in \var{sw.devices}: \func{availDev}(\var{dr},\var{c},\var{fs})$
\end{outline}
}

\full{\func{canRun} does not require \var{c} to be live, because \func{canRun} is used to check that new members of a replica set can run the replicated software, and it is sometimes useful to allow failed computers as new members, as discussed above in the paragraph ``Allowing failed new members''.}

The transition relation is defined by cases on the type of action.  The definition uses the notation $f[x := y]$ to denote the function that is the same as $f$ except that it maps $x$ to $y$.  This notation can also be applied to objects, regarded as functions from attributes to their values.  The definition ensures that if \var{cfg} is valid, then \var{cfg'} is valid.  It also ensures that functionality and device requirements of newly started software instances and replicated software instances are satisfied in failure state \var{fs}, hence their functionality is available in \var{fs}.\full{  Note that an existing resumable fast-starting software instance with persistent state may remain in the configuration if its functionality and device requirements are not met in the current configuration and failure state, and even if the computer it is running has failed, since it can potentially resume execution in the future.}

\begin{itemize}
\item $\tuple{\var{cfg}, \var{fs}} \stackrel{\func{changeReps}(\var{rsi},\var{C},\var{c})}\rightarrow 
\tuple{\var{cfg'}, \var{fs}}$
where \var{cfg'} = \var{cfg}[\var{\RSI} := $\var{\RSI} \union \{\var{rsi}[\var{members} := \var{C}]\} \setminus \var{rsi}]$ if 
\begin{outline}
        \1[$\land$] $\var{rsi} \in \var{cfg.\RSI} \land \func{resourcesOK}(\var{cfg'})$
        \1[$\land$] $\exists \var{Q} \subseteq \var{rsi.computers}: |\var{Q}| \geq \var{rsi.\repProt.reconfigQ} \land \func{live}(\var{Q}, \var{fs})$ \fmlanote{A reconfiguration quorum is live.}
        \1[$\land$] $\var{C} \subseteq \var{rsi.computers} \lor (\var{rsi.sw.fastStarting} \land (\neg \var{rsi.sw.persisState} \lor \var{rsi.sw.smallPersisState}))$ \fmlanote{New members can be added only if the software is fast-starting and the persistent state, if any, is small (hence can be copied timely to new members).}
        \1[$\land$] $\var{rsi.\repProt.active} \implies (\forall \var{c} \in \var{C} : \func{canRun}(\var{c}, \var{rsi.sw}, \var{cfg'}, \var{fs}))$ \fmlanote{For active replication, all members can run \var{rsi.sw}.}
        \1[$\land$] $\neg \var{rsi.\repProt.active} \implies \var{c} \in \var{C} \land \func{live}(\var{c}, \var{fs}) \land \func{canRun}(\var{c}, \var{rsi.sw}, \var{cfg'}, \var{fs})$ \fmlanote{For passive replication, the primary is a replica, is live, and can run \var{rsi.sw}.}
\end{outline}

\item $\tuple{\var{cfg}, \var{fs}} \stackrel{\func{stop}(\var{si})}\rightarrow
\tuple{(\var{cfg}[\var{\SI} := \var{cfg.\SI} \setminus \{\var{si}\}], \var{fs}}$ 
if $\var{si} \in \var{cfg.\SI}$

\item $\tuple{\var{cfg}, \var{fs}} \stackrel{\func{stopRep}(\var{rsi})}\rightarrow
\tuple{\var{cfg}[\var{\RSI} := \var{cfg.\RSI} \setminus \{\var{rsi}\}], \var{fs}}$ 
if $\var{rsi} \in \var{cfg.\RSI}$

\item $\tuple{\var{cfg}, \var{fs}} \stackrel{\func{start}(\var{si})}\rightarrow
\tuple{\var{cfg'}, \var{fs}}$ 
where $\var{cfg'} = \var{cfg}[\var{\SI} := \var{cfg.\SI} \union \{\var{si}\}]$ if
\begin{outline}
        \1[$\land$] $si \not\in \var{cfg.\SI} \land \func{live}(\var{si.computer}, \var{fs})$ 
        \1[$\land$] $\func{resourcesOK}(\var{cfg'})$
        \1[$\land$] \func{canRun}(\var{si.computer},\var{si.sw},\var{cfg'},\var{fs})
        \1[$\land$] $\var{si.sw.fastStarting} \land \neg \var{si.sw.persisState} \land \var{si.sw.resumable}$
        \1[$\land$] $\var{si.sw.singleInstance} \implies \var{si.sw} \not\in (\var{cfg.\SI.sw} \union \var{cfg.\RSI.sw})$
\end{outline}

\item $\tuple{\var{cfg}, \var{fs}} \stackrel{ \func{move}(\var{si},\var{c})}{\rightarrow} 
\tuple{\var{cfg'}, \var{fs}}$ 
where $\var{cfg'} = \var{cfg}[\var{\SI} := \var{cfg.\SI} \union {\var{si}[\var{computer}:=\var{c}]} \setminus \{\var{si}\}]$ if
\begin{outline}
        \1[$\land$] $\var{c} \neq \var{si.computer} \land \var{si.sw.migratable}$
        \1[$\land$] $\func{live}({\var{si.computer}, \var{c}}, \var{fs})$
        \1[$\land$] $\func{resourcesOK}(\var{cfg'}) \land \func{canRun}(\var{c}, \var{si.sw}, \var{cfg'}, \var{fs})$
        \1[$\land$] $\neg \var{si.sw.persisState} \lor \var{si.sw.smallPersisState}$
\end{outline}
\end{itemize}
}

New instances of ``stateful'' software (i.e., software $sw$ satisfying $\var{sw.persisState} \lor \neg\var{sw.resumable}$) cannot be started during reconfiguration, except that new members (i.e., computers) can be added to existing replicated instances of stateful software, provided the persistent state, if any, is small (hence can be copied quickly to new replicas).  This reflects the assumption that a software component's state must be maintained continuously from the beginning of execution.  (If the state did not need to be maintained from the beginning, then there would be no harm in losing it due to a failure and starting a fresh instance of the software, contradicting the definition of stateful software.) %
Since replication is used only for stateful software, this implies reconfiguration does not create new replicated software instances.  Relaxing these assumptions and adding a \func{startRep} action for use when appropriate is straightforward.

We define a \emph{reconfiguration relation} \func{reconfig}(\var{cfg},\var{cfg'},\var{fs}) that holds if configuration \var{cfg} with failed set \var{fs} can be reconfigured to configuration \var{cfg'}.%
\short{  The straightforward way to define this relation is to first define simpler transition relations describing the effect of each type of reconfiguration action, and then define this relation to hold if there exists a sequence of reconfiguration actions that leads from state $\tuple{\var{cfg}, \var{fs}}$ to state $\tuple{\var{cfg'}, \var{fs}}$.}%
\medfull{  The straightforward definition of this relation is that it holds if there exists a sequence of reconfiguration actions that leads from state $\tuple{\var{cfg}, \var{fs}}$ to state $\tuple{\var{cfg'}, \var{fs}}$.}  Unfortunately, this\medfull{ definition} leads to inefficient implementations that search over many sequences of actions, likely including many that lead to the same configuration.

We overcome this difficulty by giving an alternate characterization of the reconfiguration relation as a logical formula that directly expresses the possible differences between \var{cfg} and \var{cfg'}; the sequence of actions that produces those differences is implicit.  This characterization supports efficient implementation.%
\medfull{  Formally, \func{reconfig}(\var{cfg},\var{cfg'},\var{fs}) holds if:
\begin{outline}
        \1[$\land$] \func{valid}(\var{cfg'})
        \1[$\land$] $\{(\var{rsi.sw}, \var{rsi.\repProt}) \;:\; \var{rsi} \in \var{cfg'.\RSI}\} \subseteq \{(\var{rsi.sw}, \var{rsi.\repProt}) \;:\; \var{rsi} \in \var{cfg.\RSI}\}$ \fmlanote{Reconfiguration can stop but not start replicated software instances; cannot change replication protocol of existing instances.}
        \1[$\land$] $\forall \var{si'} \in \var{cfg'.\SI} \setminus \var{cfg.\SI}:$ %
            \2[$\land$] $\func{live}(\var{si'.computer}, \var{fs})$
             \2[$\land$] $\func{canRun}(\var{si'.computer},\var{si'.sw},\var{cfg'},\var{fs})$
            \2[$\land$] 
       		\3[$\lor$]
            $\var{si'.sw.fastStarting} \land \neg\var{si'.sw.persisState} \land \var{si'.sw.resumable}$ \fmlanote{This disjunct is for \emph{start}.}
 			\3[$\lor$]
            \4[$\land$] 
           $\neg\var{si'.sw.persisState} \lor \var{si'.sw.smallPersisState}$
            \4[$\land$] 
            $\var{si'.sw.migratable}$
            \4[$\land$] 
             $(\exists \var{si} \in \var{cfg.\SI} \setminus \var{cfg'.\SI}: \var{si.sw}=\var{si'.sw})$  \fmlanote{This disjunct is for \emph{move}.\full{  If \var{si.sw} is unassigned from multiple computers, we don't know which instance was moved; this is acceptable, because all instances have the same functionality.}}
        \1[$\land$] $\forall \var{rsi'} \in \var{cfg'.\RSI}:$ \fmlanote{Check membership changes.}
            \2[] let \var{rsi} = the element of \var{cfg.\RSI} with \var{rsi.sw} = \var{rsi'.sw} \fmlanote{The above subseteq condition ensures it exists.}
            \2[] $\var{rsi'.computers} \neq \var{rsi.computers} \implies {}$
                \3[$\land$] $\exists \var{Q} \subseteq \var{rsi.computers}:\\ |\var{Q}| \geq \var{rsi.\repProt.reconfigQ} \land \func{live}(\var{Q}, \var{fs})$ \fmlanote{A reconfiguration quorum is live.}
                \3[$\land$] 
            \4[$\lor$] 
                $\var{rsi'.computers} \subseteq \var{rsi.computers}$
            \4[$\lor$] 
                $\var{rsi.sw.fastStarting} \land (\neg\var{rsi.sw.persisState} \lor \var{rsi.sw.smallPersisState})$ \fmlanote{New members can be added only if the software is  fast-starting and its persistent state, if any, is small.}
                \3[$\land$] $\var{rsi.\repProt.active} \implies (\forall \var{c} \in \var{rsi'.computers} : \func{canRun}(\var{c}, \var{rsi'.sw}, \var{cfg'}, \var{fs}))$ \fmlanote{For active replication, all members can run \var{sw}.}
                \3[$\land$] $\neg \var{rsi.\repProt.active} \implies \func{live}(\var{rsi'.primary}, \var{fs}) \land \func{canRun}(\var{rsi'.primary}, \var{rsi'.sw}, \var{cfg'}, \var{fs})$ \fmlanote{For passive replication, the primary is live, and can run \var{sw}. Note that \func{valid}(\var{cfg'}) ensures $\var{rsi'.primary} \in \var{rsi'.computers}$.}
\end{outline}
}

A software instance on a failed computer is {\em dead} if its execution cannot usefully be resumed upon recovery of that computer; its execution can usefully be resumed if the software is resumable, is fast-starting, and has persistent state.
Let \func{removeDead}(\var{cfg}, \var{fs}) denote \var{cfg} with dead software instances on failed computers removed. 
A state $\tuple{\var{cfg}, \var{fs}}$ is \emph{valid} if \var{cfg} is valid and does not contain dead software instances, i.e., $\var{cfg}=\func{removeDead}(\var{cfg}, \var{fs})$.  A state $\tuple{\var{cfg}, \var{fs}}$ is an \emph{initial state} if it is valid and $\var{fs}$ is empty.
\full{Note that all states reachable from initial states by sequences of failure, recovery, and reconfiguration transitions are valid; the proof is straightforward.}

Reconfiguration enables achieving a desired level of resilience with fewer resources.  This benefit arises in a variety of situations, including the following.  (1) The software providing a functionality is not stateful and not replicated, and the computer running it can fail.
Reconfiguration can start a new instance of the software on a different computer, or a new instance of a different software providing that functionality, e.g., if the other computer has a different CPU architecture or OS, or fewer computational resources (requiring a lighter-weight software, perhaps with lower QoS).
(2) The software providing a functionality is passively replicated and one or more replicas can fail.  Reconfiguration is needed to replace the failed replicas with new ones, to tolerate subsequent failures.  (3) A majority of replicas providing an actively replicated service can fail but not simultaneously.  Reconfiguration can add new replicas after each set of simultaneous failures, to ensure a reconfiguration quorum is live when a subsequent failure occurs.\full{  For example, suppose the software is initially replicated on computers \{\var{c1},\var{c2},\var{c3}\}, and two computers can fail, but not simultaneously.  Suppose \var{c1} fails.  If the system immediately reconfigures by removing \var{c1} from the set of replicas and adding \var{c4} (possibly stopping other software on \var{c4} to meet resource constraints), it can continue to provide the service if \var{c2} subsequently fails.  If the system does not reconfigure when \var{c1} fails, the system would be stuck when \var{c2} fails: \var{c4} cannot be added at that point, due to lack of a live reconfiguration quorum.}

\mysection{Resilience and reconfiguration problem definitions}
\label{sec-prob}

This section defines resilience, along with auxiliary concepts, and then defines two core reconfiguration problems.

A {\em resilience requirement} consists of a failure model \var{fm} and a set \var{critFns} of {\em critical functionalities}, i.e., functionalities that must be continuously available in every scenario involving failures consistent with \var{fm}.  %

A failed set \var{fs} is \emph{maximal} with respect to a failure model \var{fm} if no strict superset of \var{fs} is consistent with \var{fm}; in other words, while all components in \var{fs} are failed with the specified failure types, no additional failures will happen.

Given a system model and a resilience requirement with failure model \var{fm} and critical functionalities \var{critFns}, we say that a state with configuration \var{cfg} and failed set \var{fs} is \emph{resilient}, denoted $\func{resilient}(\var{cfg}, \var{fs}, \var{fm}, \var{critFns})$, if (1) all critical functionality is available in that state, and (2) for each set of additional failures $\Delta$ consistent with \var{fm}, the system can be reconfigured after those failures to a new configuration \var{cfg'} such that (2a) all critical functionality is available in the resulting state $\tuple{\var{fs}\union\Delta, \var{cfg'}}$, and (2b) the resulting state is resilient.  The base cases of this recursive definition of resilience correspond to maximal failed sets with respect to \var{fm}.\medfull{  Formally, $\func{resilient}(\var{cfg}, \var{fs}, \var{fm}, \var{critFns})$ if
\begin{outline}
    \1[$\land$] $\func{avail}(\var{critFns},\var{cfg},\var{fs})$
    \1[$\land$] $\forall \var{fs'} \in \func{nextFS}(\var{fm}, \var{fs}): \exists \var{cfg'}$:\medium{\vspace{-1ex}}
        \2[$\land$] $\func{reconfig}(\func{removeDead}(\var{cfg}, \var{fs'}),\var{cfg'},\var{fs'})$
        \2[$\land$] $\func{resilient}(\var{cfg'}, \var{fs'}, \var{fm}, \var{critFns})$
\end{outline}
Note that $\func{removeDead}(\var{cfg}, \var{fs'})$ is used here because $\tuple{\var{cfg}, \var{fs}} \stackrel{\var{fail(fs'\setminus fs)}}\rightarrow \tuple{\func{removeDead}(\var{cfg}, \var{fs'}), \var{fs'}}$.}  Our definition of resilience does not explicitly consider recoveries.  Considering recoveries is unnecessary, because they are not guaranteed to occur and therefore cannot be relied on to provide resilience.

\begin{theorem}
Considering only worst-case failed sets in the definition of resilience, i.e., replacing \func{nextFS} with \func{nextFSworst}, leads to an equivalent definition of resilience. \label{thm-worstcase}
\end{theorem}
Informally, this holds because a system that is resilient to a set of failures occurring simultaneously is also resilient to the same failures occurring in sequence (e.g., one at a time).  The proof is straightforward.

The first core reconfiguration problem is defined as follows.  Given a system model, resilience requirement with failure model \var{fm} and critical functionalities \var{critFns}, and failed set \var{fs}, the \emph{resilient reconfiguration problem} is to find (1) one or more configurations \var{cfg} such that the state $\tuple{\var{cfg},\var{fs}}$ is resilient with respect to \var{fm} and \var{critFns} 
 and (2) a \emph{reconfiguration policy} that determines reconfiguration actions to execute from a given resilient state in response to failures and recoveries in order to reach another resilient state; or, if no such configurations exist, report this.
\full{The \emph{one-resilient reconfiguration problem} is the same as the resilient reconfiguration problem, except replace ``\func{resilient}'' with ``\func{1resilient}''.}
Typically we want to find resilient initial states, i.e., $\var{fs}=\emptyset$.

Some resilient configurations may be preferable to others, for example, because they require less hardware, use less power, or provide higher QoS.  We capture the  preference in a {\em quality metric} on configurations, which is a function from configurations to a totally ordered set.
A sample quality metric is $\func{quality}(\var{cfg}) = \tuple{\func{QoS}(\var{cfg}), \func{cost}(\var{cfg})}$, where $\func{QoS}(\var{cfg})$ is the number of instances of preferred software components (i.e., $\var{sw.preferred}=\const{true})$ running in \var{cfg}, and $\func{cost}(\var{cfg})$ is the total number of instances of software components running in it, counting a replicated software instance \var{rsi} as $|\var{rsi.computers}|$ instances.  We use lexicographic ordering on these tuples.

The second core reconfiguration problem takes configuration quality into account.  Given also a quality metric on configurations, the \emph{best resilient reconfiguration problem} is the same as the resilient reconfiguration problem, except the goal is to find one or more resilient configurations with the highest quality among resilient configurations, and a reconfiguration policy whose  reconfiguration actions lead to a highest-quality resilient state among the reachable ones.
\full{  The \emph{best one-resilient reconfiguration problem} is defined similarly.} %

\mysection{Algorithm}
\label{sec:algo}

A straightforward implementation of our framework that performs a naive search for resilient configurations or best resilient configurations is about 140 lines of DistAlgo code, ignoring imports, type declarations, and trivial code for initialization, logging, and printing.  Sections \ref{sec-irrelevant} and \ref{sec-reduction} below describe two significant optimizations to the naive search.\medfull{  Section \ref{sec-search} presents the overall algorithm.}  We also implemented an optimized algorithm for \func{nextFSworst}, which avoids generating and then dropping non-worst-case failed sets in some cases.
These optimizations add about 190 lines, ignoring trivial code for initialization, logging, etc.

\subsection{Optimization: Avoid irrelevant configurations} 
\label{sec-irrelevant}

Our algorithm starts by generating two sets of configurations: a set \var{initCfg} of initial configurations to use for the top-level search for a resilient \short{one}\medium{initial configuration}, and a set \var{allCfg} of valid configurations to use repeatedly when searching for the new configuration $\var{cfg'}$ in case (2b) of the definition of resilience.

Our algorithm for generating \var{allCfg} is optimized to avoid enumerating invalid configurations as well as valid configurations that are \emph{irrelevant} in the sense that they do not affect the answers to the resilient reconfiguration problem, and they do not affect the answers to the best resilient reconfiguration problem provided the configuration quality metric does not prefer expensive (high-resource) configurations.  A configuration is irrelevant if it (1) lacks software instances that provide critical functionalities, (2) contains too few replicas in replicated software instances (relative to the number of possible simultaneous failures and the quorum sizes of the replication algorithm), so that it clearly does not satisfy the conditions on \var{cfg'} in the definition of resilience, or (3) has too many software instances, or too many replicas in replicated software instances, so that it clearly uses more resources than necessary. 
\short{  A detailed definition of irrelevant configuration is in \cite{adaptiveDist-arxiv}.}\medfull{

Our algorithm uses the following criteria to identify irrelevant configurations.  Given a set $\var{critFns}$ of critical functionality, a software component \var{sw} is \emph{critical} if it is the only software in the system that provides some functionality in $\var{critFns}$.  A critical software component \var{sw} requires replication if it has persistent state, is not fast-starting, or is not resumable.  A configuration is \emph{irrelevant} if it contains:
\begin{itemize}
  \item no instances of one or more critical software components; 
  \item  a replicated instance \var{rsi} of some critical software component \var{sw} that requires replication, and \var{rsi} has (a) fewer than 3 or more than $2f+1$ replicas if $\var{rsi.repAlg.progressQ}=\const{majority}$ or $\var{rsi.repAlg.reconfigQ}=\const{majority}$,
or (b) more than $f+1$ replicas otherwise, where $f$ is the maximum number of computers that can fail simultaneously, according to the failure model; or
    \item multiple instances (replicated or unreplicated) of any software component that can be used remotely.
\end{itemize}}

Our algorithm for generating \var{initCfg} uses a broader definition of irrelevant configurations that extends the above definition with the following clauses.
(1) For replicated software instance \var{rsi} with a passive replication protocol, if all computers in \var{rsi.computers} are \emph{equivalent} (i.e., have the same values for all attributes listed in Section \ref{sec-system}), then one is chosen (arbitrarily) as the primary, and configurations that differ only by having a different primary for \var{rsi} are irrelevant.  
(2) A configuration \var{cfg} is irrelevant if it contains a replicated software instance \var{rsi} whose replica set \var{rsi.computers} contains a computer\short{ that cannot run \var{rsi.sw} in configuration \var{cfg} due to unsatisfied dependence, compatibility, capacity, or distribution constraints.}\medfull{ $c$ not satisfying \func{canRun}(\var{c},\var{rsi.sw},\var{cfg},$\emptyset$)}.

\subsection{Optimization: \rsence quotient reduction}
\label{sec-reduction}

Quotient reduction \cite{emerson2000} is a technique for reducing the number of states of a system that need to be considered during verification by model checking of all possible states. It is by exploiting an equivalence relation on states.  The equivalence relation must be designed to preserve the system behavior and the properties being verified.  This ensures that it is sufficient to consider one representative of each equivalence class when verifying those properties.  Symmetry reduction is a well-known type of quotient reduction, where the equivalence relation is based on a symmetry relation; clause (1) in the extension of the definition of irrelevant reconfigurations at the end of Section~\ref{sec-irrelevant} for generating \var{initCfg} is a simple example of a symmetry reduction.
 
We developed a quotient reduction based on a novel equivalence relation called \emph{relocatable-software equivalence (\rsence)}.  A software component is relocatable (defined formally below) if it can be relocated during reconfiguration by starting a new instance on a different computer.  The main idea is that two configurations are \rsent if they differ only in the computers on which relocatable software components are running.
Intuitively, if a relocatable software component can be run on either of two computers, then we can arbitrarily choose to run it on one of them, and ignore the possibility of running it on the other, because if the chosen computer fails, we can start a new instance of that software component on the other computer during reconfiguration.

\rsence applies to systems where the dependency relation between functionalities is acyclic, with the dependency relation defined by: functionality \var{fn_1} depends on functionality \var{fn_2} if some software component that provides \var{fn_1} requires \var{fn_2}\medfull{, i.e., $\exists \var{sw}\in\var{sys.sw}: \var{sw.fn}=\var{fn_1} \land \var{fn_2} \in \var{sw.reqFns}$}.  This applicability condition is typically satisfied in practice, because software systems typically do not have cyclic dependencies between services.
Generalizing \rsence to allow cyclic dependencies is a direction for future work.

A software component \var{sw} is \emph{relocatable} in configuration \var{cfg}, denoted \reloc(\var{sw}, \var{cfg}), if:\short{ (1) $sw$ is resumable, does not have persistent state, and is fast-starting; (2) if any software component in \var{cfg} depends on \var{sw}'s functionality, then \var{sw} supports remote use; and (3) software components in \var{cfg} on whose functionality \var{sw} depends support remote use.}\medfull{
\begin{outline}
	\1[$\land$] $\var{sw.fastStarting} \land \neg \var{sw.persisState} \land \var{sw.resumable}$
    \1[$\land$] $(\exists \var{si}\in \var{cfg.\SI}: \var{sw.fn}\in\var{si.sw.fnReq}) \implies \var{sw.remoteUse}$ \fmlanote{If another software instance depends on \var{sw}, then \var{sw} supports remote use.}
    \1[$\land$] $\forall \var{si} \in (\var{cfg.\SI} \union \var{cfg.\RSI}): \var{si.sw.fn} \in \var{sw.fnReq} \implies \var{si.sw.remoteUse}$ \fmlanote{Software instances on which \var{sw} depends support remote use.}
\end{outline}}
The first condition ensures that, during reconfiguration, the software can be stopped and then re-started on a different computer.  The second and third conditions ensure that software dependencies cannot prevent relocation of this software during reconfiguration.

Configurations \var{cfg_1} and \var{cfg_2} are \emph{\rsent} if they are valid and, letting 
$R_i = \{\var{si} \in \var{cfg_i.\SI} \;:\; \reloc(\var{si.sw}, \var{cfg_i})\}$, there exists a bijection $f$ from $R_1$ to $R_2$\medfull{ (with the interpretation that \var{si_1} in $R_1$ is relocated from \var{si_1.computer} in \var{cfg_1} to \var{f(si_1).computer} in \var{cfg_2} if those two computers are different, and \var{si_1} stays on the same computer otherwise)} such that:
\begin{outline}
    \1[\medfull{$\land$}] $\forall \var{si_1} \in R_1: 
    \var{si_1.sw} = \var{f(si_1).sw} \land{}\\
    \var{si_1.sw.devices} \,\intersect\, \var{si_1.computer.devices} = \var{si_1.sw.devices} \,\intersect\, \var{f(si_1).computer.devices}$\medfull{ \fmlanote{Each relocatable software instance is running in the two configurations on computers with the same types of integrated devices among the device types it uses.}}
    \1[$\land$] $\var{cfg_1.\SI} \setminus R_1 = \var{cfg_2.\SI} \setminus R_2$ \fmlanote{The configurations contain the same instances of non-relocatable software components.}
    \1[$\land$] $\var{cfg_1.\RSI} = \var{cfg_2.\RSI}$\medfull{ \fmlanote{The configurations contain the same replicated software instances.}}
\end{outline}
Note that the computers on which a relocatable software component \var{sw} is running in \var{cfg_1} and \var{cfg_2} do not need to be identical.\short{  They do need to have the same types of integrated devices used by \var{sw}; otherwise failure of a stand-alone device could break the equivalence, i.e., cause \var{sw} to be runnable on one of those computers but not the other.}\medfull{  The requirement on integrated devices ensures that failure of a stand-alone device cannot break the equivalence by causing \var{sw} to be runnable on one of the computers but not the other.}

\begin{lemma}
If configurations \var{cfg_1} and \var{cfg_2} are \rsent, then for all failure states \var{fs} such that $\tuple{\var{cfg_1}, \var{fs}}$ and $\tuple{\var{cfg_2}, \var{fs}}$ are valid, the same functionality is available in these states\full{, i.e., $F_1=F_2$ where 
$F_i = \setc{\var{fn}}{\func{avail}(\var{fn},\var{cfg_i},\var{fs})}$}.
\label{lemma-func}
\end{lemma}\vspace{-2.5ex}
\myparagraph{Proof sketch}
\full{We start with a proof sketch, to build intuition, and then a detailed proof.}
The proof is by induction on the set of available functionalities in each configuration, partially ordered by the dependency relation between functionalities.\medfull{  This is why \rsence applies to systems with an acyclic dependency relation.}  
The proof is straightforward\medfull{ using %
the definitions of relocatable software and \rsent}.
\qed

\begin{lemma}
\rsence is an equivalence relation.
\label{lemma-equiv}
\end{lemma}\vspace{-2.5ex}
\myparagraph{Proof sketch} The proof is straightforward.
\qed

\begin{theorem}
\rsence preserves resilience.  Precisely, for every failure model \var{fm}, set of critical functionalities \var{critFns}, failed set \var{fs} consistent with \var{fm}, and valid configurations \var{cfg_1} and \var{cfg_2}, 
if states $\tuple{\var{cfg_1},\var{fs}}$ and $\tuple{\var{cfg_2},\var{fs}}$ are valid, 
and \var{cfg_1} and \var{cfg_2} are \rsent,
then
\func{resilient}(\linebreak[0]\var{cfg_1},\var{fs},\var{fm},\var{critFns}) iff
\func{resilient}(\var{cfg_2},\var{fs},\var{fm},\var{critFns}).
\label{thm-preserves}
\end{theorem}\vspace{-2.5ex}
\myparagraph{Proof sketch}
We show that resilience of \var{cfg_1} implies resilience of \var{cfg_2}; the opposite implication follows from the same reasoning, due to the symmetry in the statement of the theorem.  In short, we need to show that (1) all functionality in \var{critFns} is available in \var{cfg_2}, and (2) for each failure state \var{fs'} in \func{nextFS}(\var{fm},\var{fs}), there exists a resilient configuration $\var{cfg'}$ to which \func{removeDead}(\var{cfg_2},\var{fs'}) can reconfigure.  Conclusion (1) follows from the assumption that \var{cfg_1} is resilient and Lemma \ref{lemma-func}.
Consider conclusion (2).  Resilience of \var{cfg_1} implies that, for each failure state \var{fs'} in \func{nextFS}(\var{fm},\var{fs}), there exists a resilient configuration \var{cfg'} to which \func{removeDead}(\var{cfg_1},\var{fs'}) can reconfigure, using some sequence \var{S} of reconfiguration actions.  Also, it is reasonably straightforward to show, using the premises of the theorem and the definition of \rsence, that \func{removeDead}(\var{cfg_2},\var{fs'}) can reconfigure to \func{removeDead}(\var{cfg_1},\var{fs'}), by using stop and start actions to relocate relocatable software instances that are on different computers in \func{removeDead}(\var{cfg_1},\var{fs'}) and \func{removeDead}(\var{cfg_2},\var{fs'}). \func{removeDead}(\var{cfg_1},\var{fs'}) can then reconfigure to \var{cfg'} using actions \var{S}.
\qed

\subsection{Optimized search-based algorithm}
\label{sec-search}

Our search-based algorithm for finding resilient configurations, incorporating the above optimizations, works as follows:
\begin{enumerate}
    \item Generate the sets \var{initCfg} and \var{allCfg} of all relevant initial configurations and all relevant configurations, respectively.
    \item Compute equivalence classes with respect to \rsence for \var{allCfg}.  Specifically, create a dictionary \var{D} that maps each configuration \var{cfg} to a configuration that uniquely represents \var{cfg}'s equivalence class.
    \item Create a set \var{initCfg'} containing the representatives of the equivalence classes of configurations in \var{initCfg}.
    \item For each configuration \var{cfg} in \var{initCfg'}, check $\func{resilient}(\var{cfg}, \emptyset, \var{fm}, \var{critFns})$\medfull{, where \var{fm} is the given failure model and \var{critFns} is the set of critical functionalities}.
\end{enumerate}

To evaluate \func{resilient}(\var{cfg},\var{fs},\var{fm},\var{critFns}), first check that all critical functionalities are available in state $\tuple{\var{cfg},\var{fs}}$,\medfull{ i.e., that $\func{avail}(\var{critFns},\tuple{\var{cfg},\var{fs}})$ holds,} 
and then for each failed set \var{fs'} in \func{nextFSworst}(\var{fm},\var{fs}) (recall that these are the subset-maximal sets of failures that can occur in a state with failed set $\var{fs}$):
    \begin{enumerate}
        \item Compute the set, called \var{next}, of next configurations by iterating over \var{allCfg} and adding each configuration \var{cfg_1} such that \func{reconfig}(\func{removeDead}(\var{cfg},\var{fs'}),\var{cfg_1},\var{fs'}) and \var{next} does not already contain a member of  \var{cfg_1}'s equivalence class.
        \item Check whether \func{resilient}(\var{cfg_1},\var{fs'},\var{fm},\var{critFns}) holds for some \var{cfg_1} in \var{next}.  If not, return \const{false}.
    \end{enumerate}
If this loop exits normally (without returning \const{false}), return \const{true}.  The recursion terminates when\medfull{ the maximum number of failures allowed by the failure model have occurred, and} \func{nextFSworst}(\var{fm},\var{fs}) returns an empty set.\short{  The algorithm can easily be extended to compute and store the sequence of reconfiguration actions needed to transition from a state to a resilient successor state; this provides the reconfiguration policy.}

\medfull{The algorithm can be extended easily (and with little additional runtime cost) to compute the sequence of reconfiguration actions needed to transition from a state to a resilient successor state, if one exists.}
To find resilient states that optimize a given configuration quality metric, the algorithm sorts \var{initCfg'} in descending order by configuration quality, and checks resilience of the configurations in that order. With the extension to store reconfiguration actions, in the algorithm for evaluating \var{resilient}, among the configurations \var{cfg_1}  found to be resilient in step 2, we store the sequence of reconfiguration actions that leads to a configuration of highest quality configuration among those.  A straightforward extension is to add a notion of reconfiguration cost (e.g., based on reconfiguration time) and store the lowest-cost sequence of reconfiguration actions from \var{cfg} to \var{cfg_1}.

\medfull{\section{SMT-based approach}
\label{sec-smt}

We also implemented much of our framework in Z3~\cite{z3smt2008}, a state-of-the-art SMT solver, using Z3's Python API.  We chose Z3 because we expected that our mathematical definitions could be translated easily into its logic, and that its sophisticated symbolic reasoning engine would provide good scalability.  However, we encountered multiple difficulties.

First, translating our definitions into its logic was difficult.  Z3 is based on first-order logic and does not support quantification over sets or functions.  Since the definition of resilience quantifies over failed sets and configurations, those concepts cannot be represented in a straightforward way as sets or functions, respectively.  Instead, we needed to encode them (and other information) as collections of Boolean variables.  This greatly complicates the code and significantly increases the size of the generated Z3 formulas.  This, in turn, made debugging the implementation, and achieving high assurance in its correctness, difficult.

Second, the recursive definition of resilience cannot be expressed directly in Z3's logic, which does not support recursion.  The recursion would need to be simulated in the Python driver program, using a loop that repeatedly calls Z3.  

Third, the best resilient reconfiguration problem can be supported easily in Z3 only for simple quality metrics for which the problem can be reduced to MaxSAT.  Supporting more general quality metrics, including the sample quality metric in Section \ref{sec-prob}, would be difficult.

Due to these difficulties, our implementation of the SMT-based approach is incomplete and is not used in the experiments in Section \ref{sec-eval}.
}

\mysection{Evaluation}
\label{sec-eval}

We evaluated our approach on a model of an autonomous driving system.  The software architecture models \short{\href{https://github.com/autowarefoundation/autoware}{Autoware}}\medfull{Autoware~\cite{autoware}},
a prominent open-source software stack for self-driving vehicles.  The computing platform models QualComm's
\short{\href{https://www.qualcomm.com/products/application/automotive/autonomous-driving}{Snapdragon Ride} platform}\medfull{Snapdragon Ride platform \cite{snapdragonride}},
which uses multiple smaller embedded computers with ARM processors, for greater scalability and power-efficiency than a single large computer. 

The hardware model includes several sensors---camera, GPS, IMU, LIDAR, and radar---and two or more quad-core ARM-based embedded computers running Safe RT Linux, a real-time OS.  The embedded computers have enough RAM that it is not a limiting factor.  We assume the network is also real-time,  hence the system is synchronous.

We modeled five main software components with the following functionalities:\short{ (1) perception: performs object detection, tracking, and prediction using data from radar, LIDAR, and camera; (2) localization: estimates the location of the vehicle using data from GPS, IMU, camera, etc.; (3) planning: produces a trajectory towards a provided goal, starting from the current position and velocity; (4) control: generates commands (for steering, braking, etc.) that will make the vehicle follow a given trajectory; (5) vehicle interface: interfaces the autonomous driving software with a specific vehicle model.}\medfull{
\begin{itemize}
\item perception: performs object detection, tracking, and prediction using data from radar, LIDAR, and camera;
\item localization: estimates the location of the vehicle using data from GPS, IMU, camera, etc.;
\item planning: produces a trajectory towards a provided goal, starting from the current position and velocity; 
\item control: generates commands (for steering, braking, etc.) that will make the vehicle follow a given trajectory; 
\item vehicle interface: interfaces the autonomous driving software with a specific vehicle model.
 \end{itemize}}

The system includes preferred Safe RT Linux versions of all five software components.  Each software component requires one core\unused{ and 1 GB RAM}, except the perception component, which requires two cores\unused{ and 2 GB RAM}.  Perception maintains critical volatile state, which is needed for object tracking and prediction, and hence this software component is not resumable.  The other four software components are resumable.  All of the software components are fast-starting.  None maintain critical persistent state.  The primary-backup protocol is used for replication, since the system is synchronous, hardware resources are limited, and primary-backup requires fewer replicas to tolerate a given number of failures.  We consider one type of failure: crashes of embedded computers.

We applied our algorithm to two versions of this system with different sets of hardware and software components, described below.  We report the time needed to find \emph{all} resilient initial configurations---or, when \rsence reduction is used, to find all equivalence classes of resilient initial configurations---and then sort them by quality, using the sample configuration quality metric in Section
\ref{sec-prob}.  This comprehensive result is not required by the problem definitions in Section \ref{sec-prob} but forces comprehensive exploration of the search space and eliminates potential variation in running time due to non-determinism in iteration orders, which can cause the first resilient configuration to appear earlier or later in the search.  All experiments were run on a MacBook Air M1 (2020) with 8 GB RAM running macOS 11.6, Python 3.7.12 and DistAlgo 1.1.0b15, with garbage collection disabled for more consistent running times.   We validated consistency of running times by repeating experiments three times and noting that all measured running times were within 5\% of the mean.

\subsection{Example 1: Failover to laptop}

In this example, the set of hardware is extended with a laptop running Linux.  It is not normally part of the autonomous driving system but can be used for some autonomous driving functionality in case of failures. The set of software is extended with non-preferred Linux versions of the planning and control components.  Localization and perception are less suitable for migration to the laptop, because they process streams of sensor data and hence require high-bandwidth connections to the vehicle's network.  The vehicle interface component is by nature hardware-specific and requires direct connection to the vehicle's network.  All five functionalities are critical.

We illustrate the results of the resilience analysis by describing one highest-quality resilient initial configuration in a system with two embedded computers, \var{c0} and \var{c1}, and how it reconfigures after a failure.  In the initial configuration, unreplicated instances of the control and vehicle interface software run on \var{c0}, unreplicated instances of the planning and localization software run on \var{c1}, and a replicated instance of the perception software runs on \var{c0} (the primary) and~\var{c1}. 

Upon failure of \var{c0}, the system reconfigures by stopping the instance of planning software on \var{c1}, starting an instance of the vehicle interface software on \var{c1}, starting the Linux versions of the control and planning software on the laptop, making \var{c1} the primary replica for the perception software, and removing \var{c0} as a replica for the perception software (we continue to represent it as a replicated software instance, even though there is only one replica, so we can easily add a replica if \var{c0} recovers).

\newcommand{\ncomp}{N_C}
\newcommand{\nfail}{N_F}

Table \ref{tab-cfgs}-(left) presents the numbers of configurations and the numbers of equivalence classes of configurations generated.  The first column reports $\ncomp$, the number of embedded computers, and $\nfail$, the maximum number of failures of embedded computers, which is always $\ncomp-1$; also, we always use $\var{maxSimult}=\nfail$.  The second and third columns report the sizes of \var{allCfg} and \var{initCfg}, respectively.  The remaining columns report the numbers of equivalence classes of configurations in \var{allCfg}, \var{initCfg}, and \var{resilCfg}, respectively, where \var{resilCfg} is the set of resilient initial configurations found.

Table \ref{tab-times}-(left) presents the running times of three versions of the algorithm: no Q.R., using no quotient reduction; partial Q.R., using partial \rsence quotient reduction (i.e., for \var{initCfg} only, in step 3 of the pseudocode for search in Section \ref{sec-search}); and full Q.R., using full \rsence quotient reduction (i.e., for \var{initCfg} and \var{next}).  The running time is partitioned into generate time (the time taken to enumerate all relevant configurations, and to generate equivalence classes based on \rsence if applicable) and analyze time (the rest  of the running time).  ``T-O'' (time-out) indicates the experiment was aborted after 1 hour.

We see that partial Q.R.\ drastically reduces the running time for systems with more computers, e.g., by a factor of 30 for Example 2 with three computers, and even larger factors with more computers (though the exact speedups in these cases were not measured due to the 1-hour time-out).  Even with this optimization, the running time grows quickly with the number of computers, indicating that there is opportunity for additional symmetry reductions, beyond the simple one used in the definition of irrelevant configurations for \var{initCfg}.

We also observe that the algorithm with full Q.R. has comparable running time as the algorithm with partial Q.R.  This shows that the time saved by exploring fewer states is balanced by the overhead of applying the reduction to \var{next} at every step.  
The full reduction further reduces the number of explored configurations, and therefore is still beneficial if RAM is at a premium.

\begin{table}
        \setlength{\tabcolsep}{0.6\tabcolsep}
        \begin{tabular}{|c|r|r|r|r|r|}
             \hline
             $\langle\ncomp,$ &\multicolumn{2}{c|}{\# of configs}&\multicolumn{3}{c|}{\# of equiv.\ classes}\\\cline{2-6}
             ~$\nfail\rangle$ & \var{allCfg} & \var{initCfg} & \var{allCfg} & \var{initCfg} & \var{resilCfg} \\
             \hline
             $\tuple{2, 1}$ & 256 & 204 & 36 & 27 & 9 \\
             $\tuple{3, 2}$ & 4,332 & 2,607 & 108 & 63 & 9 \\
             $\tuple{4, 3}$ & 36,992 & 17,648 & 288 & 135 & 9 \\
             $\tuple{5, 4}$ & 224,720 & 88,015 & 720 & 279 & 9 \\
             \hline
        \end{tabular}
        \ieeeonly{\\\\\\}
        \begin{tabular}{|c|r|r|r|r|r|}
             \hline
             $\langle\ncomp,$ &\multicolumn{2}{c|}{\# of configs}&\multicolumn{3}{c|}{\# of equiv.\ classes}\\\cline{2-6}
             ~$\nfail\rangle$ & \var{allCfg} & \var{initCfg} & \var{allCfg} & \var{initCfg} & \var{resilCfg} \\
             \hline
             $\tuple{2, 1}$ & 162 & 128 & 24 & 18 & 6 \\
             $\tuple{3, 2}$ & 2,520 & 1,512 & 72 & 42 & 6 \\
             $\tuple{4, 3}$ & 20,720 & 9,872 & 192 & 90 & 6 \\
             $\tuple{5, 4}$ & 123,120 & 48,190 & 480 & 186 & 6 \\
             \hline
        \end{tabular}
        \full{%
        \\\\\\
        \begin{tabular}{|c|r|r|r|r|r|}
             \hline
             \multirow{2}{*}{Parameters}&\multicolumn{2}{c|}{Length}&\multicolumn{3}{c|}{Number of equivalence classes}\\\cline{2-6}
             & allConfigs & initConfigs & allConfigs & initConfigs  & resilConfigs \\
             \hline
             (2, 1) & 678 & 525 & 72 & 54 & 18 \\
             (3, 2) & 23640 & 14018 & 432 & 252 & 36 \\
             (4, 3) & 410176 & 194052 & 2304 & 1080 & T-O \\
             (5, 4) & 5038080 & 1963176 & T-O & T-O & T-O \\
             \hline
        \end{tabular}
        }
	 \short{\\}
    \caption{Numbers of configurations analyzed for Examples 1 (left)\shortmed{ and 2 (right)}\full{, 2 (middle) and 3 (bottom)}.}\short{\vspace{-1ex}}
    \label{tab-cfgs}
\end{table}

\newcommand{\generate}{\,gen.}
\begin{table}
        \setlength{\tabcolsep}{0.6\tabcolsep}
        \begin{tabular}{|c|r|r|r|r|r|r|}
             \hline
             $\langle\ncomp,$ &\multicolumn{2}{c|}{no Q.R.}&\multicolumn{2}{c|}{partial Q.R.}&\multicolumn{2}{c|}{full Q.R.}\\\cline{2-7}
             ~$\nfail\rangle$ & \generate & analyze & \generate & analyze & \generate & analyze\\
             \hline
             $\tuple{2, 1}$ & 0.1 & 2.0 & 0.1 & 0.3 & 0.1 & 0.1 \\
             $\tuple{3, 2}$ & 1.4 & 171.1 & 1.4 & 9.8 & 1.6 & 6.1 \\
             $\tuple{4, 3}$ & 10.8 & T-O & 11.6 & 110.2 & 13.3 & 106.7 \\
             $\tuple{5, 4}$ & 69.7 & T-O & 67.3 & 1,263.2 & 84.5 & 1,283.4 \\
             \hline
        \end{tabular}
        \ieeeonly{\\\\\\}
        \begin{tabular}{|c|r|r|r|r|r|r|}
             \hline
             $\langle\ncomp,$ &\multicolumn{2}{c|}{no Q.R.}&\multicolumn{2}{c|}{partial Q.R.}&\multicolumn{2}{c|}{full Q.R.}\\\cline{2-7}
             ~$\nfail\rangle$ & \generate & analyze & \generate & analyze & \generate & analyze \\
             \hline
             $\tuple{2, 1}$ & 0.1 & 0.6 & 0.1 & 0.1 & 0.1 & 0.1 \\
             $\tuple{3, 2}$ & 0.5 & 74.4 & 0.6 & 2.0 & 0.6 & 1.9 \\
             $\tuple{4, 3}$ & 4.5 & T-O & 4.51 & 29.4 & 5 & 29.2 \\
             $\tuple{5, 4}$ & 24.5 & T-O & 26 & 367.8 & 31.8 & 347.4 \\
             \hline
        \end{tabular}
        \full{%
        \\\\\\
        \begin{tabular}{|p{1cm}|p{1cm}|p{1cm}|p{1cm}|p{1cm}|p{1cm}|p{1cm}|}
             \hline
             \multirow{2}{*}{Parameters}&\multicolumn{2}{p{2cm}|}{No Q.R.}&\multicolumn{2}{p{2cm}|}{partial Q.R.}&\multicolumn{2}{p{2cm}|}{full Q.R.}\\\cline{2-7}
             & gen. & analyze & gen. & analyze & gen. & analyze \\
             \hline
             $\tuple{2, 1}$ & 0.36 & 13.7 & 0.34 & 1.34 & 0.39 & 1.47 \\
             $\tuple{3, 2}$ & 11.78 & T-O & 14.83 & 179.79 & 18.59 & 171.67 \\
             $\tuple{4, 3}$ & 237.73 & T-O & 287.68 & T-O & 505.78 & T-O \\
             $\tuple{5, 4}$ & 2103.71 & T-O & T-O & T-O & T-O & T-O \\
             \hline
        \end{tabular}
        }
        \short{\\}
    \caption{Running times, in seconds, for Examples 1 (left)\shortmed{ and 2 (right)}\full{, 2 (middle) and 3 (bottom)}.  ``gen.'' abbreviates ``generate''.}\short{\vspace{-4ex}}
    \label{tab-times}
\end{table}

\subsection{Example 2: Failover to smartphone-guided human operator}
\label{sec-example-phone}

In this example, there is a human operator who can take over some driving functions in case an embedded computer fails.  The set of hardware is extended with an Android smartphone, and the set of software is extended with a non-preferred manual-control software component that runs on Android smartphones.  
The manual-control software uses the smartphone's display and speaker to convey instructions for the human operator to drive the vehicle along the path computed by the planning component, in a similar way as Google Maps.
When the manual-control software is used for control, the vehicle interface component is not needed; therefore, vehicle interface functionality is not critical in this example.  The other four functionalities are critical.

After a failure, the system reconfigures in a similar way as in Example 1, by starting an instance of the manual-control software on the smartphone, starting any needed instances of other software components on the surviving embedded computer, etc.

Performance results for Example 2 appear in Tables \ref{tab-cfgs}-(right) and \ref{tab-times}-(right).  The performance observations for Example 1 apply to Example 2 too.  We also note that resilience analysis for Example 2 explores fewer configurations and is faster than for Example 1, because Example 2 involves only one software component with a non-preferred alternative version, while Example 1 involves two.

\mysection{Conclusion}
\label{sec:conclusion}

We developed a general automated framework for making systems resilient by optimized adaptive distribution and replication of software components, including a model-finding algorithm that finds resilient configurations and reconfiguration policies using state-space exploration with multiple optimizations, including a quotient reduction based on a novel equivalence relation.

\short{D}\medfull{Interesting d}irections for future work include: 
modeling and analysis of network failures and partitions; 
probabilistic failure models and probabilistic resilience properties;
more detailed timing analysis to consider response-time requirements, failover latency, etc.; 
partially-ordered quality metrics that reflect trade-offs between cost and resilience;
new performance optimizations, including symmetry reductions that reduce the set of explored failed sets; as well as new case studies, especially involving multi-agent systems such as collections of robots or drones, with functionalities replicated across agents. %

\articleonly{\paragraph{Acknowledgments.} \thanksText }

\ieeeonly{\bibliographystyle{IEEEtran}}\acmonly{\bibliographystyle{ACM-Reference-Format}}\lncsonly{\bibliographystyle{splncs04}}\articleonly{\bibliographystyle{alpha}}
\bibliography{references}

 \end{document}